\documentclass[aps,prd,twocolumn,nofootinbib]{revtex4}

\usepackage{graphicx}
\usepackage{hyperref}
\usepackage{slashed}
\usepackage{color}

\begin{document}

\title {Heavy $P$-wave quarkonium production via Higgs decays}

\author{Qi-Li Liao$^{1}$}
\email{xiaosueer@163.com}
\author{Ya Deng$^{1}$}
\email{dengya2007@163.com}
\author{Yan Yu$^{1}$}
\email{yanyu2002034@163.com}
\author{Guang-Chuan Wang$^{1}$}
\author{Guo-Ya Xie$^{2}$}
\address{$^{1}$College of Mobile Telecommunications Chongqing  University of Posts and Telecom, Chongqing 401520, P.R. China\\$^{2}$ Chongqing  University of Posts and Telecom, Chongqing 400065, P.R. China}

\date{\today}
\begin{abstract}
The production of the heavy quarkonium, i.e., $|(c\bar{b})[n]\rangle$ (or $|(b\bar{c})[n]\rangle$), $|(c\bar{c})[n]\rangle$, and $|(b\bar{b})[n]\rangle$- quarkonium [$|(Q\bar{Q'})[n]\rangle$-quarkonium for short], through Higgs $H^{0}$ boson semiexclusive decays is evaluated within the NRQCD framework, where $[n]$ stands for the production of the two color-singlet $S$-wave states, $|(Q\bar{Q'})[^1S_0]_{\textbf{1}} \rangle$ and $|(Q\bar{Q'})[^3S_1]_{\textbf{1}} \rangle$, the production of the four color-singlet $P$-wave states, i.e., $|(Q\bar{Q'})[^1P_0]_{\textbf{1}}\rangle$, $|(Q\bar{Q'})[^3P_J]_{\textbf{1}}\rangle$ (with $J =[0, 1, 2]$). Moreover, according to the velocity scaling rule of the NRQCD, the production of the two color-octet components, $|(Q\bar{Q'})g[^1S_0]_{\textbf{8}} \rangle$ and $|(Q\bar{Q'})g[^3S_1]_{\textbf{8}} \rangle$, are also taken into account. The ``improved trace technology" to derive the simplified analytic expressions at the amplitude level is adopted, which shall be useful for dealing with these decay channels. If all higher heavy quarkonium states decay completely to the ground states, it should be obtained  $\Gamma{(H^0\to |(c\bar{b})[^1S_0]_{\textbf{1}}\rangle)}=15.14$ KeV, $\Gamma{(H^0\to |(c\bar{c})[^1S_0]_{\textbf{1}}\rangle)}=1.547$ KeV, and $\Gamma{(H^0\to |(b\bar{b})[^1S_0]_{\textbf{1}}\rangle)}=1.311$ KeV. The production of $5.6\times10^{5}$ Bc meson, $4.7\times10^{4}$ charmonium meson, and $4.9\times10^{4}$ bottomonium meson per year in Higgs decays at the HE/HL-LHC can be obtained.\\


\end{abstract}

\maketitle

\section{Introduction}

Since the Higgs boson of the standard model (SM) has been found by CMS \cite{cms1} and ATLAS \cite{atlas1} at the Large Hadron Collider (LHC) in July 2012. The lots of experimental results and review papers on the Higgs boson production and decay were obtained by the CMS and ATLAS at the LHC \cite{km,cmsatlas,ck}. With the growingly accumulated date, the properties of a new particle are consistent with those of Higgs boson predicted by SM \cite{cms2,atlas2}. Though the LHC offers obvious advantages in proving very high energy and very large rates in typical reactions, the measuring precision will be restricted due to the complicated background.

The most precise measurements will be performed in the clean environment of the future electron-positron collider for the proposed Higgs factory, like the International Linear Collider (ILC) \cite{hb} and the Circular Electron-Positron Collider (CEPC) \cite{cepc}. It is well known that the main production processes of the Higgs boson in electron-positron collider collisions are the Higgs-strahlung process $e^{+}e^{-}\rightarrow H^{0}Z^{0}$ and the $W^{+}W^{-}$ fusion process $e^{+}e^{-}\rightarrow \nu_{e}\bar{\nu_{e}}H^{0}$. The cross section for the Higgs-strahlung process is dominant at the low energy. For $\sqrt{s}=500$ GeV, the cross section for the $W^{+}W^{-}$ fusion is dominant. The cross section for the $Z^{0}Z^{0}$ fusion process $e^{+}e^{-}\rightarrow e^{+}e^{-}H^{0}$ increases significantly with the center-of-mass (c.m.) energy increasing, and can exceeds that of $Z^{0}H^{0}$ production around $\sqrt{s} = 1$ TeV. These processes can be well used to test the Higgs-gauge couplings. The Higgs self-coupling can be studied through the double Higgs boson production processes $e^{+}e^{-}\rightarrow Z^{0}H^{0}H^{0}$ and $e^{+}e^{-}\rightarrow \nu_{e}\bar{\nu_{e}}H^{0}H^{0}$ at the ILC. The absolute values of the Higgs coupling to bosons, gluons and heavy fermions can also be measured. When updated to the Super Proton-Proton Collider (SPPC), researchers can even measure the Higgs self-coupling, which is regarded as the holy grail of experimental particle physics high luminosity/energy (HL/HE-LHC) scenarios are designed for the LHC \cite{eml,lhc}. Running at center-of-mass energy $\sqrt{s} = 14$ TeV, cross-section of the Higgs boson production at the LHC is about $55$ pb (gluon-gluon fusion process dominates). Given that the integrated luminosity is 3 $ab^{-1}$, the HL-LHC would produce $1.65\times10^{8}$ Higgs events \cite{lhc}. While at the HE-LHC who runs at $\sqrt{s} = 33$ TeV, the cross-section of the Higgs boson production would be about $200$ pb, hence the Higgs boson events per year can be obtained $6.0\times10^{8}$.

With the above mentioned excellent platforms, rare Higgs boson decay processes, like the heavy quarkonium production in the Higgs boson decays, might be observed for the first time. Pioneer investigation on the search of $H^0\longrightarrow J/\Psi \gamma$ and $H^0\longrightarrow \Upsilon(nS) \gamma$ has been carried out by ATLAS \cite{ga}. Theoretically, some related calculations have been done \cite{vkaa,ad,nv,cfk,gf}. Within the nonrelativistic quantum chromodynamics (NRQCD) formulism \cite{nrqcd} and light-cone methods \cite{gs}, both direct and indirect production mechanism and relativistic corrections to $H^0\longrightarrow J/\Psi \gamma$ and $H^0\longrightarrow \Upsilon(nS) \gamma$ are studied \cite{gf}. In Ref. \cite{jjcq}, the $B_c$ ($B^*_c$) meson production via Higgs boson decays under the NRQCD \cite{nrqcd} is systematic investigated. Where both the quantum chromodynamics (QCD) and the quantum electrodynamics (QED) contributions are included. It is found that the production of $B_c$ ($B^*_c$) meson through the QED contributions is very smaller than through the QCD, for example, $\Gamma(H^0\to |(b\bar{c})[n]\rangle+\bar{b}c)_{QED}/\Gamma(H^0\to |(b\bar{c})[n]\rangle+\bar{b}c)_{QCD}\sim 10^{-5}$. In comparison with the QCD one, QED contribution is negligible for production the heavy quarkonium through the Higgs boson decays. So it is only studied the QCD contribution for the Higgs boson decays production the heavy quarkonium in this paper.

The LHCb, ATLAS, and CMS Collaboration experiments have published studies of the $B_c$ meson production and of the double $J/\Psi$ production \cite{r,cc,ga}. Since its discovery by the CDF Collaboration \cite{fa}, the $B_c$ meson being the unique ``doubly heavy-flavored" meson in the SM has aroused great interest. The direct hadronic production of the $B_c$ meson has been studied systematically in Refs. \cite{kar,spb,cyg,aam,cjx,ccpx}. Therefore, investigation of the heavy quarkonium production through $H^0$ decays is worthwhile and meaningful. The heavy quarkonium is presumed to be a nonrelativistic bound state of the heavy quark and antiquark. The study of the heavy quarkonium, e.g., $|(b\bar{c})[n] \rangle$ (or $|(c\bar{b})[n] \rangle$), $|(c\bar{c})[n] \rangle$, and $|(b\bar{b})[n] \rangle$-quarkonium, can help us to achieve a deeper understanding of the QCD in both the perturbative and nonperturbative sectors. A very practical theoretical tool to deal with the processes involving heavy quarkonium is the NRQCD \cite{nrqcd}, in which the low-energy interactions are organized by the expansion in $v$, where $v$ stands for the typical relative velocity of the heavy quark and antiquark inside of the heavy quarkonium. The heavy quarkonium production itself is very useful for high precision physics in the electroweak sector and testing the perturbative QCD (pQCD) \cite{nb1,nb2,glb}. For compensation, it would be helpful to study its indirect production mechanisms. A systematic study of the heavy quarkonium production through $W^{\pm}$, $Z^0$ boson, and $t$ (or $\bar{t}$) quark decays can be found in the literature \cite{vbkw,ekt,cck,w,lx,tbc2,zbc0,zbc1,zbc2,wbc1,wbc2}.

Due to a high collision energy and high luminosity at the HL/HE-LHC, sizable amounts of the heavy quarkonium events can be produced through $H^0$ decays \cite{eml,lhc}. So these channels may be an important supplement for other measurements at the HL/HE-LHC. In this work, we will study the $|(c\bar{b})[n]\rangle$ (or $|(b\bar{c})[n]\rangle$), $|(c\bar{c})[n]\rangle$ and $|(b\bar{b})[n]\rangle$-quarkonium production in Higgs boson decays under the NRQCD, where $[n]$ stands for $1^1S_0$, $1^3S_1$, $1^1P_0$, $n^3P_J$ ($J=[0, 1, 2]$). To deal with heavy quarkonium production through $H^0$ semiexclusive decays, one needs to derive the pQCD calculable squared amplitudes. But the analytical expression for the usual squared amplitude $|\Sigma|^2$ becomes too complex and lengthy for more (massive) particles in the final states and for higher-level Fock states to be generated for the emergence of massive-fermion lines in the Feynman diagrams, especially to derive the amplitudes of the $P$-wave states. To solve the problem, the ``improved trace technology" is suggested and developed in the literature \cite{lx,tbc2,zbc0,zbc1,zbc2,wbc1,wbc2}; it deals with the process directly at the amplitude level. We will continue to adopt improved trace technology to derive the analytical expression for all the above-mentioned decay channels.

The rest of the paper is organized as follows. We introduce the calculation techniques for the $H^0$ boson semiexclusive decays to $|(Q\bar{Q'})[n]\rangle$-quarkonium within the NRQCD formulism in Sec.II. In Sec.III, we calculate the production of $|(c\bar{b})[n]\rangle$ (or $|(b\bar{c})[n]\rangle$), $|(c\bar{c})[n]\rangle$, and $|(b\bar{b})[n]\rangle$-quarkonium through $H^0$ decay channels, i.e., $H^0\to |(c\bar{b})[n]\rangle+\bar{c}b$, $H^0\to |(c\bar{c})[n]\rangle+\bar{c}c$, and $H^0\to |(b\bar{b})[n]\rangle+\bar{b}b$, with new parameters \cite{lx} for the $|(Q\bar{Q'})[n]\rangle$-quarkonium, an estimation of events at the HL/HE-LHC. Then we make some discussions on the theoretical uncertainties of the decays widths by the masses of the $|(c\bar{b})[n]\rangle$-quarkonium. The final section is reserved for a summary.

\section{CALCULATION TECHNIQUES AND FORMULATION}

The $H^0$ boson decays semiexclusive processes for the heavy quarkonium production can be analogous dealt with, i.e.,$H^0\to |(c\bar{b})[n]\rangle+\bar{c}b$ (or $H^0\to |(b\bar{c})[n]\rangle+\bar{b}c$), $H^0\to |(c\bar{c})[n]\rangle+\bar{c}c$, and $H^0\to |(b\bar{b})[n]\rangle+\bar{b}b$. According to the NRQCD factorization formula \cite{nrqcd}, the square of the semiexclusive amplitude can be written as the production of the perturbatively calculable short-distance coefficients and the nonperturbative long-distance factors, the so-called nonperturbative NRQCD matrix elements. Its total decay widths $d\Gamma$ can be factorized as

\begin{equation}
d\Gamma=\sum_{n} d\hat\Gamma(H^0\to |(Q\bar{Q'})[n]\rangle+ \bar{Q'}Q) \frac{\langle{\cal O}^H(n) \rangle}{N_{col}},
\end{equation}
where $N_{col}$ refers to the number of colors, $n$ stands for the involved state of  $|(Q\bar{Q'})[n]\rangle$-quarkonium. $N_{col}=1$ for singlets and $N_{col}=N^2-1$ for octets; $\langle{\cal O}^{H}(n)\rangle$ is the nonperturbative matrix element which describes the hadronization of a $Q\bar{Q'}$ pair into the observable quark hadron state and is proportional to the transition probability of the perturbative state $Q\bar{Q'}$ into the bound state $|(Q\bar{Q'})[n]\rangle$. As for the color-singlet components, the nonperturbative matrix elements can be directly related either to the wave functions at the origin for $S$-wave states or the first derivative of the wave functions at the origin for $P$-wave states \cite{nrqcd}, which can be computed via the potential NRQCD (pNRQCD) \cite{pnrqcd1, nb2, yellow} and/or lattice QCD \cite{lat1} and/or the potential models \cite{lx, pot1, pot2, pot3, pot4, pot5}. Although we do not know the exact values of the two decay color-octet matrix elements, $|(Q\bar{Q'})g[^1S_0]_{\textbf{8}} \rangle$ and $|(Q\bar{Q'})g[^3S_1]_{\textbf{8}} \rangle$, we know that they are one order in $v^2$ higher than the $S$-wave color-singlet matrix elements according to NRQCD scale rule. More specifically, based on the velocity scale rule, we have:

\begin{eqnarray}
&&\langle{(Q\bar{Q'}})g[^1S_0]_{\textbf{8}}|{\cal O}[^1S_{0}]_{\textbf{8}}| {(Q\bar{Q'}})g[^1S_0]_{\textbf{8}}\rangle \simeq \nonumber\\
&&~~~\Delta_{S}(v^2) \cdot \langle{(Q\bar{Q'}})[^1S_0]_{\textbf{1}}|{\cal O}[^1S_0]_{\textbf{1}}| {(Q\bar{Q'}})[^1S_0] _{\textbf{1}}\rangle ,\nonumber\\
&&\langle{(Q\bar{Q'}})g[^3S_1]_{\textbf{8}}|{\cal O}[^3S_1]_{\textbf{8}}| {(Q\bar{Q'}})g[^3S_1]_{\textbf{8}}\rangle \simeq \nonumber\\
&&~~~\Delta_{S}(v^2) \cdot \langle{(Q\bar{Q'}})[^3S_1]_{\textbf{1}}|{\cal O}[^3S_1]_{\textbf{1}}| {(Q\bar{Q'}})[^3S_1]_{\textbf{1}}\rangle.
\end{eqnarray}
Where the second equation comes from the vacuum-saturation approximation. The thickened subscripts of the $(Q\bar{Q'})$ denote for color indices, $1$ for color singlet and $8$ for color-octet; the relevant angular momentum quantum numbers are shown in the parentheses accordingly. Here $v$ is the relative velocity between the components, $\bigtriangleup_s({v^2})$ is of the order $v^2$ or so, and we take it to be within the region of $0.10$, $0.20$, $0.30$ for $|(b\bar{b})g[^1S_0]_{\textbf{8}}\rangle$ ($|(b\bar{b})g[^3S_1]_{\textbf{8}}\rangle$), $|(b\bar{c})g[^1S_0]_{\textbf{8}} \rangle$ ($|(b\bar{c})g[^3S_1]_{\textbf{8}}\rangle$), and $|(c\bar{c})g[^1S_0]_{\textbf{8}} \rangle$ ($|(c\bar{c})g[^3S_1]_{\textbf{8}} \rangle$), respectively; which is in consistent with the identification: $\bigtriangleup_s{v^2}\sim\alpha_s(Mv)$ and has covered the possible variation due to the different ways to obtain the wave functions at the origin ($S$-wave) and the first derivative of the wave functions at the origin ($P$-wave).

The short-distance decay width $\hat\Gamma(H^0)$ can be expressed as
\begin{equation}
d\hat\Gamma(H^0\to |(Q\bar{Q'})[n]\rangle+\bar{Q}Q')= \frac{1}{2 m_{M}} \overline{\sum}  |M(n)|^{2} d\Phi_3,
\end{equation}
where $\overline{\sum}$ means that one needs to average over the spin states of the initial particle and to sum over the color and spin of all the final particles. In the $H^0$ rest frame, the three-particle phase space can be written as
\begin{equation}
d{\Phi_3}=(2\pi)^4 \delta^{4}\left(k_{0} - \sum_f^3 q_{f}\right)\prod_{f=1}^3 \frac{d^3{\vec{q}_f}}{(2\pi)^3 2q_f^0}.
\end{equation}

\begin{figure}
\includegraphics[width=0.40\textwidth]{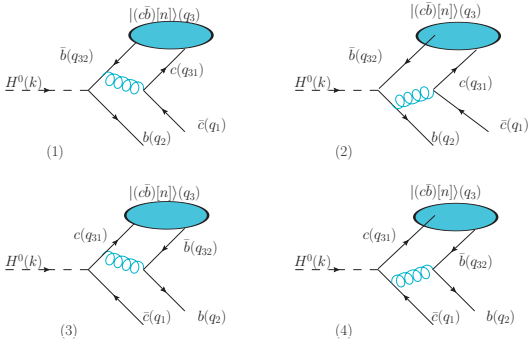}
\caption{ Feynman diagrams for the process $H^0(k)\rightarrow |(c\bar{b})[n]\rangle(q_3) + b(q_2)+\bar{c}(q_1)$, where $|(c\bar{b})[n]\rangle$ stands for $|(c\bar{b})[^1S_0]_{\textbf{1}}\rangle$, $|(c\bar{b})[^3S_1]_{\textbf{1}}\rangle$, $|(c\bar{b})[^1P_1]_{\textbf{1}}\rangle$, $|(c\bar{b})[1^3P_J]_{\textbf{1}}\rangle$ (with $J=[0 , 1, 2]$), $|(c\bar{b})g[^1S_0]_{\textbf{8}}\rangle$ and $|(c\bar{b})g[1^3S_1]_{\textbf{8}}\rangle$ quarkonium, respectively.} \label{feyn1}
\end{figure}

The process to simplify the $1 \to 3$ phase space with a massive quark/antiqark in the final state has been dealt with in greater detail in Refs.~\cite{tbc2,zbc1}. To shorten the paper, we shall not present it here, but the interested reader may turn to these references for the detailed technology. With the help of the formulas listed in Refs.~\cite{tbc2,zbc1}, one can not only derive the whole decay widths but also obtain the corresponding differential decay widths that are helpful for experimental studies, such as $d\Gamma/ds_{1}$, $d\Gamma/ds_{2}$, $d\Gamma/d\cos\theta_{12}$, and $d\Gamma/d\cos\theta_{23}$, where $s_{1}=(q_1+q_2)^2$, $s_{2}=(q_1+q_3)^2$, $\theta_{12}$ is the angle between $\vec{q}_1$ and $\vec{q}_2$, and $\theta_{23}$ between $\vec{q}_2$ and $\vec{q}_3$.

To better illustrate the Feynman diagrams of the above three processes as $H^0(k) \rightarrow |(Q\bar{Q'})[n]\rangle(q_3)+Q'(q_2)+\bar{Q}(q_1)$,  the Feynman diagrams of the process $H^0(k)\rightarrow |(c\bar{b})[n]\rangle(q_3) + b(q_2)+\bar{c}(q_1)$ is presented in Fig. \ref{feyn1} for example, where the intermediate gluon should be hard enough to produce a $b\bar{b}$ pair or $c\bar{c}$ pair, so the amplitude is pQCD calculable.

These amplitudes can be generally expressed as
\begin{equation} \label{amplitude}
i{\cal{M} }= {\cal{C}} {\bar {u}_{s i}}({q_2}) \sum\limits_{n = 1}^{m} {{\cal A} _n } {v_{s' j}}({q_1}),
\end{equation}
where $m$ stands for the number of Feynman diagrams, $s$ and $s'$ are spin states, and $i$ and $j$ are color indices for the outing $Q$-quark and $\bar{Q}$-quark, respectively. The overall factor $\cal{C}=\cal{C}_S$ stands for the specified quarkonium in the color-singlet, where ${\cal C}_S={C_{F}~g~g_s^2 \cdot \delta_{ij}}/({2 m_{H^{0}}~\sqrt{3}})$. Where $\cal{C}=\cal{C}_O$ stands for the color-octet states, where ${\cal C}_O={g~g_s^2 \cdot (\sqrt{2} T^aT^bT^a)_{ij}}/({2 m_{H^{0}}~\sqrt{2}})$, here $\sqrt{2}T^a$ stands for the color factor of the color-octet quarkonium. ${\cal A} _n $'s in the formulas are the amplitudes of $H^0(k) \rightarrow |(Q\bar{Q'})[n]\rangle(q_3) +Q'(q_2)+ \bar{Q}(q_1)$.

By using the improved trace technology, one can sequentially obtain the squared amplitudes, and the numerical efficiency can also be greatly improved ~\cite{lx,tbc2,zbc0,zbc1,zbc2,wbc1,wbc2}. We adopt the ``improved trace technology" to simplify the amplitudes $M_{ss^{\prime}}$ at the amplitude level for the above-mentioned processes, the amplitudes ${\cal A}_n$ of $H^0(k) \rightarrow |(Q\bar{Q'})[n]\rangle(q_3) +Q'(q_2)+ \bar{Q}(q_1)$ for the $S$-wave states are

\begin{eqnarray}
{\cal A}_1 &=& \left[{m_Q}{\gamma_\alpha} \frac{\Pi^{0(\nu)}_{q_3}(q)}{(q_2 + {q_{32}})^2} {\gamma_\alpha} \frac{(\slashed{q}_2+\slashed{q}_3)+{m_Q}}{(q_2+q_3 )^2 - m_Q^2} \right]_{q=0}, \label{A1}\\
{\cal A}_2 &=& \left[{m_Q}{\gamma_\alpha} \frac{\Pi^{0(\nu)}_{q_3}(q)}{(q_2 + {q_{32}})^2} \frac{-(\slashed{k}-\slashed{q}_{31})+{m_Q}}{(k-q_{31} )^2 - m_Q^2}{\gamma_\alpha} \right]_{q=0}, \label{A2}\\
{\cal A}_3 &=& \left[{m_{Q'}} \frac{-(\slashed{q_1}+\slashed{q}_{3})+{m_{Q'}}}{(q_1+q_{3} )^2 - m_{Q'}^2}{\gamma_\alpha} \frac{\Pi^{0(\nu)}_{q_3}(q)}{(q_1 + {q_{31}})^2} {\gamma_\alpha} \right]_{q=0}, \label{A3}\\
{\cal A}_4 &=& \left[{m_{Q'}} {\gamma_\alpha} \frac{-(\slashed{k}-\slashed{q}_{32})+{m_{Q'}}}{(k-q_{32} )^2 - m_{Q'}^2} \frac{\Pi^{0(\nu)}_{q_3}(q)}{(q_1 + {q_{31}})^2} {\gamma_\alpha} \right]_{q=0}. \label{A4}
\end{eqnarray}

For the $^1P_1$-wave states, ${\cal A}_n$ can be written as
\begin{widetext}
\begin{eqnarray}
{\cal A}^{S=0,L=1}_1 &=& \varepsilon_l^{\mu}(q_3) \frac{d}{dq_\mu} \left[{m_Q}{\gamma_\alpha} \frac{\Pi^{0}_{q_3}(q)}{(q_2 + {q_{32}})^2} {\gamma_\alpha} \frac{(\slashed{q}_2+\slashed{q}_3)+{m_Q}}{(q_2+q_3 )^2 - m_Q^2} \right]_{q=0}, \label{A5}\\
{\cal A}^{S=0,L=1}_2 &=& \varepsilon_l^{\mu}(q_3) \frac{d}{dq_\mu} \left[{m_Q}{\gamma_\alpha} \frac{\Pi^{0}_{q_3}(q)}{(q_2 + {q_{32}})^2} \frac{-(\slashed{k}-\slashed{q}_{31})+{m_Q}}{(k-q_{31} )^2 - m_Q^2}{\gamma_\alpha} \right]_{q=0}, \label{A6}\\
{\cal A}^{S=0,L=1}_3 &=& \varepsilon_l^{\mu}(q_3) \frac{d}{dq_\mu} \left[{m_{Q'}} \frac{-(\slashed{q_1}+\slashed{q}_{3})+{m_{Q'}}}{(q_1+q_{3} )^2 - m_{Q'}^2}{\gamma_\alpha} \frac{\Pi^{0(\nu)}_{q_3}(q)}{(q_1 + {q_{31}})^2} {\gamma_\alpha} \right]_{q=0}, \label{A7}\\
{\cal A}^{S=0,L=1}_4 &=& \varepsilon_l^{\mu}(q_3) \frac{d}{dq_\mu} \left[{m_{Q'}} {\gamma_\alpha} \frac{-(\slashed{k}-\slashed{q}_{32})+{m_{Q'}}}{(k-q_{32} )^2 - m_{Q'}^2} \frac{\Pi^{0}_{q_3}(q)}{(q_1 + {q_{31}})^2} {\gamma_\alpha} \right]_{q=0}. \label{A8}
\end{eqnarray}
and the $^3P_J$-wave states ($J=0,1,2$)
\begin{eqnarray}
{\cal A}^{S=1,L=1}_1 &=& \varepsilon^{J}_{\mu\nu}(q_3) \frac{d}{dq_\mu}  \left[{m_Q}{\gamma_\alpha} \frac{\Pi^{\nu}_{q_3}(q)}{(q_2 + {q_{32}})^2} {\gamma_\alpha} \frac{(\slashed{q}_2+\slashed{q}_3)+{m_Q}}{(q_2+q_3 )^2 - m_Q^2} \right]_{q=0}, \label{A9}\\
{\cal A}^{S=1,L=1}_2 &=& \varepsilon^{J}_{\mu\nu}(q_3) \frac{d}{dq_\mu} \left[{m_Q}{\gamma_\alpha} \frac{\Pi^{\nu}_{q_3}(q)}{(q_2 + {q_{32}})^2} \frac{-(\slashed{k}-\slashed{q}_{31})+{m_Q}}{(k-q_{31} )^2 - m_Q^2}{\gamma_\alpha} \right]_{q=0}, \label{A10}\\
{\cal A}^{S=1,L=1}_3 &=& \varepsilon^{J}_{\mu\nu}(q_3) \frac{d}{dq_\mu}  \left[{m_{Q'}} \frac{-(\slashed{q_1}+\slashed{q}_{3})+{m_{Q'}}}{(q_1+q_{3} )^2 - m_{Q'}^2}{\gamma_\alpha} \frac{\Pi^{\nu}_{q_3}(q)}{(q_1 + {q_{31}})^2} {\gamma_\alpha} \right]_{q=0}, \label{A11}\\
{\cal A}^{S=1,L=1}_4 &=& \varepsilon^{J}_{\mu\nu}(q_3) \frac{d}{dq_\mu} \left[{m_{Q'}} {\gamma_\alpha} \frac{-(\slashed{k}-\slashed{q}_{32})+{m_{Q'}}}{(k-q_{32} )^2 - m_{Q'}^2} \frac{\Pi^{\nu}_{q_3}(q)}{(q_1 + {q_{31}})^2} {\gamma_\alpha} \right]_{q=0}. \label{A12}
\end{eqnarray}
\end{widetext}
Here $\varepsilon_{s}(q_3)$ and $\varepsilon_{l}(q_3)$ are the polarization vectors relating to the spin and the orbit angular momentum of the $|(Q\bar{Q'})[n]\rangle$-quarkonium, $\varepsilon^{J}_{\mu\nu}(q_3)$ is the polarization tensor for the spin triplet $P$-wave states (with $J=[0 , 1, 2]$). The covariant form of the projectors can be conveniently written as
\begin{equation}
\Pi^0_{q_3}(q)=\frac{-\sqrt{m_{Q\bar{Q'}}}}{4{m_Q}{m_{Q'}}}(\slashed{q}_{32}- m_{Q'}) \gamma_5 (\slashed{q}_{31} + m_Q)\bigotimes\frac{\mathbf{1_c}}{\sqrt{N_c}},
\end{equation}
and
\begin{equation}
\Pi^\nu_{q_3}(q)=\frac{-\sqrt{m_{Q\bar{Q'}}}}{4{m_Q}{m_{Q'}}}(\slashed{q}_{32}- m_{Q'}) \gamma_\nu (\slashed{q}_{31} + m_Q)\bigotimes\frac{\mathbf{1_c}}{\sqrt{N_c}}.
\end{equation}
Here $\mathbf{1_c}$ stands for the unit color matrix with $N_c=3$ for the QCD; $q$ stands for the relative momentum between the two constituent quarks in the $|(Q\bar{Q'})[n]\rangle$-quarkonium. $q_{31}$ and $q_{32}$ are the
momenta of the two constituent quarks, i.e.,
\begin{equation}
{q_{31}}=\frac{{m_b}}{m_{Q\bar{Q'}}}q_3+q,~~~~~~~~{q_{32}}=\frac{{m_Q}}{m_{Q\bar{Q'}}}q_3-q.
\end{equation}
where $m_{Q\bar{Q'}}=m_Q+m_{Q'}$ is implicitly adopted to ensure the gauge invariance of the hard scattering amplitude.

Finally, the decay widths over $s_{1}$ and $s_{2}$ can be expressed as
\begin{equation}
d\Gamma= \frac{3}{256 \pi^3 m^3_H}( \overline{\sum}|M|^{2}) \frac{\langle{\cal O}^H(n) \rangle}{N_{col}} ds_{1}ds_{2},
\end{equation}
where $m_H$ is the mass of the $H^0$ boson, the extra factor $3$ in the numerator comes from the sum of the $Q$-quark color. The color-singlet nonperturbative matrix element $\langle{\cal O}^H(n) \rangle$ can be related either to the Schr${\rm \ddot{o}}$dinger wave function $\psi_{(Q\bar{Q'})}(0)$ at the origin for the $S$-wave quarkonium states or the first derivative of the wave function $\psi^\prime_{(Q\bar{Q'})}(0)$ at the origin for the $P$-wave quarkonium states:
\begin{eqnarray}
\langle{\cal O}^H(1S) \rangle &\simeq& |\psi_{\mid(Q\bar{Q'})[1S]\rangle}(0)|^2,\nonumber\\
\langle{\cal O}^H(1P) \rangle &\simeq& |\psi^\prime_{\mid(Q\bar{Q'})[1P]\rangle}(0)|^2.
\end{eqnarray}

As the spin-splitting effects are small, the difference between the wave function parameters for the spin-singlet and spin-triplet states at the same level are not distinguished. The Schr${\rm \ddot{o}}$dinger wave function at the origin $\Psi_{|Q\bar{Q'})[1S]\rangle}(0)$ and the first derivative of the Schr${\rm \ddot{o}}$dinger wave function at the origin $\Psi^{'}_{|(Q\bar{Q'})[1P]\rangle}(0)$ are related to the radial wave function at the origin $R_{|(Q\bar{Q'})[1S]\rangle}(0)$ and the first derivative of the radial wave function at the origin $R^{'}_{|(Q\bar{Q'})[1P]\rangle}(0)$, respectively~\cite{nrqcd,lx}.
\begin{eqnarray}
\Psi_{|(Q\bar{Q'})[1S]\rangle}(0)&=&\sqrt{{1}/{4\pi}}R_{|(Q\bar{Q'})[1S]\rangle}(0),\nonumber\\
\Psi'_{|(Q\bar{Q'})[1P]\rangle}(0)&=&\sqrt{{3}/{4\pi}}R'_{|(Q\bar{Q'})[1P]\rangle}(0).
\end{eqnarray}

The radial wave function at the origin $R_{|(Q\bar{Q'})[1S]\rangle}(0)$ and the first derivative of the radial wave function at the origin $R^{'}_{|(Q\bar{Q'})[1P]\rangle}(0)$ relate to the number of active flavor quarks $n_f$, the constituent quark mass of the $|(Q\bar{Q'})[n]\rangle$-quarkonium, and the concrete potential models, respectively~\cite{lx}. Thus, $R_{|(Q\bar{Q'})[1S]\rangle}(0)$ and $R^{'}_{|(Q\bar{Q})[1P]\rangle}(0)$ in this paper are adopted in Ref.~\cite{lx}.

\section{Numerical results and discussions}

\subsection{Input parameters}

The input parameters are adopted as the following values \cite{wtd,pdg}: $m_H =125.7$ GeV, the Higgs $H^0$ total decay with $\Gamma(H^0)=4.2$ MeV is adopted~\cite{sh}. $m_{W}=80.399$GeV, $\theta_W=\arcsin\sqrt{0.23119}$ is the Weinberg angle. We set the renormalization scale to be $m_{(c\bar{c})}$ and $m_{(c\bar{b})}$ of $|(c\bar{c})\rangle$ and $|(c\bar{b})\rangle$-quarkonium for leading-order $\alpha_s$ running , which leads to $\alpha_s=0.26$ and $m_{(b\bar{b})}$ of $|(b\bar{b})\rangle$-quarkonium for $\alpha_s=0.18$. To ensure the gauge invariance of the hard amplitude, we set the $|(Q\bar{Q'})[n]\rangle$-quarkonium mass $M$ to be $m_Q+m_{Q'}$. We adopt the values derived in Refs.~\cite{lx,pdg} and list them in Table~\ref{tabrpa}, since it is noted that the Buchm${\rm \ddot{u}}$ller and Tye potential (B.T. potential) has the correct two-loop short-distance behavior in QCD~\cite{pot2,wgs} and the decay widths are related to the constituent quark mass of the $|(Q\bar{Q}')[n]\rangle$-quarkonium.

\begin{table}
\caption{Mass of the constituent quark and radial wave functions at the origin under the B.T. potential \cite{lx}.}
\begin{tabular}{|c||c|c|c|c|}
\hline\hline
$|(Q\bar{Q'})[n]\rangle$&$|(c\bar{c})[n]\rangle$&$|(c\bar{b})[n]\rangle$&$|(b\bar{b})[n]\rangle$\\
\hline
$m_{S} ({GeV})$&~~1.48~~&~~1.45/~4.85~&~~4.71~\\
\hline
$|R_{|[1S]\rangle}(0)|^2({GeV}^3)$&~2.458~&~3.848~~&~~16.12~\\
\hline
$m_{P} ({GeV})$&~1.75~&~~1.75/4.93~~&~4.94~\\
\hline
$|R'_{|[1P]\rangle}(0)|^2({GeV}^5)$&~0.322~&~~0.518~~&~5.874~\\
\hline\hline
\end{tabular}
\label{tabrpa}
\end{table}

\subsection{Heavy quarkonium production via $H^0$ decays}

The decay widths for the $|(c\bar{b})[n]\rangle$ (or $|(b\bar{c})[n]\rangle$), $|(c\bar{c})[n]\rangle$, and $|(b\bar{b})[n]\rangle$-quarkonium states and the production channels through $H^0$ decays, i.e., $H^0\rightarrow |(c\bar{b})[n]\rangle+\bar{c}b$ (or $H^0\rightarrow |(b\bar{c})[n]\rangle+\bar{b}c$), $H^0\rightarrow |(c\bar{c})[n]\rangle+\bar{c}c$, and $H^0\rightarrow |(b\bar{b})[n]\rangle+\bar{b}b$, are listed in Tables \ref{tabrpb}, \ref{tabrpc}, and \ref{tabrpd} within the B.T. potential\cite{lx}, where the Higgs total decay with is adopted $\Gamma_{H}=4.20MeV$ \cite{sh}. If the input parameters of the Ref.\cite{jjcq} is adopted. The results are consistent with the results of this paper for $S$-wave states. As the choice of the new model parameters \cite{lx}, the calculation results of our paper and Ref.\cite{jjcq} is different. The main difference of the calculation results comes from the matrix elements $\langle{\cal O}^{H}(n)\rangle$ (wave function). But $\Gamma{(H^0\to B^{*}_c)}/ \Gamma{(H^0\to B_c)}=2.08KeV/1.53KeV=1.364$ in Ref.\cite{jjcq} is almost equal to $\Gamma{(H^0\to B^{*}_c)}/\Gamma{(H^0\to B_c)}=7.857KeV/5.736KeV=1.370$ in our paper.

\begin{table}
\caption{Decay widths and branching fractions for the production of the $|(c\bar{b})[n]\rangle$-quarkonium through Higgs boson semiexclusive decays within the B.T. potential ($n_f=3$)~\cite{lx, pot2}.}
\begin{tabular}{|c|c|c|c|}
\hline
$H^0\to |(c\bar{b})[n]\rangle+\bar{c}b$&$\Gamma{(H^0\to|(c\bar{b})[n]\rangle)}$(KeV)&$\frac{\Gamma{(H^0\rightarrow |(c\bar{b})[n]\rangle)}}{\Gamma_{H}}$\\
\hline\hline
$H^0\to |(c\bar{b})[^1S_0]_{\textbf{1}}\rangle +\bar{c}b$&~~5.736~~~&1.37$\times10^{-3}$~\\
\hline
$H^0\to |(c\bar{b})[^3S_1]_{\textbf{1}}\rangle +\bar{c}b$&~~7.857~~~&1.87$\times10^{-3}$\\
\hline
$H^0\to |(c\bar{b})[^1P_1]_{\textbf{1}}\rangle +\bar{c}b$&~~0.2761~~~&6.57$\times10^{-5}$\\
\hline
$H^0\to |(c\bar{b})[^3P_0]_{\textbf{1}}\rangle +\bar{c}b$&~~0.1838~~~&4.38$\times10^{-5}$\\
\hline
$H^0\to |(c\bar{b})[^3P_1]_{\textbf{1}}\rangle +\bar{c}b$&~~0.6706~~~&1.60$\times10^{-4}$\\
\hline
$H^0\to |(c\bar{b})[^3P_2]_{\textbf{1}}\rangle +\bar{c}b$&~~0.3521~~~&8.38$\times10^{-5}$\\
\hline
$H^0\rightarrow |(c\bar{b})g[^1S_0]_{\textbf{8}}\rangle +\bar{c}b$&~0.7170$v^{4}$~&1.71$v^{4}\times10^{-4}$\\
\hline
$H^0\to |(c\bar{b})g[^3S_1]_{\textbf{8}}\rangle +\bar{c}b$&~0.9821$v^{4}$~&2.24$v^{4}\times10^{-4}$\\
\hline\hline
\end{tabular}
\label{tabrpb}
\end{table}

\begin{table}
\caption{Decay widths and branching fractions for the production of the $|(c\bar{c})[n]\rangle$-quarkonium through Higgs boson semiexclusive decays within the B.T. potential ($n_f=3$)~\cite{lx, pot2}.}
\begin{tabular}{|c|c|c|c|}
\hline
~$H^0\rightarrow |(c\bar{c})[n]\rangle+\bar{c}b$~&~$\Gamma{(H^0\rightarrow |(c\bar{c})[n]\rangle)}$(eV)~&~$\frac{\Gamma{(H^0\rightarrow |(c\bar{c})[n]\rangle)}}{\Gamma_{H}}$~\\
\hline\hline
$H^0\rightarrow |(c\bar{c})[^1S_0]_{\textbf{1}}\rangle +\bar{c}c$&~~646.6~~~&1.54$\times10^{-4}$\\
\hline
$H^0\rightarrow |(c\bar{c})[^3S_1]_{\textbf{1}}\rangle +\bar{c}c$&~~623.7~~~&1.49$\times10^{-4}$\\
\hline
$H^0\rightarrow |(c\bar{c})[^1P_1]_{\textbf{1}}\rangle +\bar{c}c$&~~69.89~~~&1.66$\times10^{-5}$\\
\hline
$H^0\rightarrow |(c\bar{c})[^3P_0]_{\textbf{1}}\rangle +\bar{c}c$&~~101.9~~~&2.43$\times10^{-5}$\\
\hline
$H^0\rightarrow |(c\bar{c})[^3P_1]_{\textbf{1}}\rangle +\bar{c}c$&~~44.37~~~&1.06$\times10^{-5}$\\
\hline
$H^0\rightarrow |(c\bar{c})[^3P_2]_{\textbf{1}}\rangle +\bar{c}c$&~~46.02~~~&1.10$\times10^{-5}$\\
\hline
$H^0\rightarrow |(c\bar{c})g[^1S_0]_{\textbf{8}}\rangle +\bar{c}c$&80.82$v^{4}$&1.92$v^{4}\times10^{-5}$\\
\hline
$H^0\rightarrow |(c\bar{c})g[^3S_1]_{\textbf{8}}\rangle +\bar{c}c$&77.96$v^{4}$&1.86$v^{4}\times10^{-5}$\\
\hline\hline
\end{tabular}
\label{tabrpc}
\end{table}

\begin{table}
\caption{Decay widths and branching fractions for the production of the $|(b\bar{b})[n]\rangle$-quarkonium through Higgs boson semiexclusive decays within the B.T. potential ($n_f=4$)~\cite{lx, pot2}.}
\begin{tabular}{|c|c|c|c|}
\hline
~$H^0\rightarrow |(b\bar{b})[n]\rangle+\bar{c}b$~&~$\Gamma{(H^0\rightarrow |(b\bar{b})[n]\rangle)}$(eV)~&~$\frac{\Gamma{(H^0\rightarrow |(b\bar{b})[n]\rangle)}}{\Gamma_{H}}$~\\
\hline\hline
$H^0\rightarrow |(b\bar{b})[^1S_0]_{\textbf{1}}\rangle +\bar{b}b$&~~680.6~~~&1.62$\times10^{-4}$\\
\hline
$H^0\rightarrow |(b\bar{b})[^3S_1]_{\textbf{1}}\rangle +\bar{b}b$&~~513.4~~~&1.22$\times10^{-4}$\\
\hline
$H^0\rightarrow |(b\bar{b})[^1P_1]_{\textbf{1}}\rangle +\bar{b}b$&~~21.36~~~&5.09$\times10^{-6}$\\
\hline
$H^0\rightarrow |(b\bar{b})[^3P_0]_{\textbf{1}}\rangle +\bar{b}b$&~~39.83~~~&9.48$\times10^{-6}$\\
\hline
$H^0\rightarrow |(b\bar{b})[^3P_1]_{\textbf{1}}\rangle +\bar{b}b$&~38.73~~~&9.22$\times10^{-6}$\\
\hline
$H^0\rightarrow |(b\bar{b})[^3P_2]_{\textbf{1}}\rangle +\bar{b}b$&~~15.63~~~&3.72$\times10^{-6}$\\
\hline
$H^0\rightarrow |(b\bar{b})g[^1S_0]_{\textbf{8}}\rangle +\bar{b}b$&85.08$v^{4}$&2.03$v^{4}\times10^{-5}$\\
\hline
$H^0\rightarrow |(b\bar{b})g[^3S_1]_{\textbf{8}}\rangle +\bar{b}b$&64.17$v^{4}$&1.53$v^{4}\times10^{-5}$\\
\hline\hline
\end{tabular}
\label{tabrpd}
\end{table}

\begin{figure}
\includegraphics[width=0.40\textwidth]{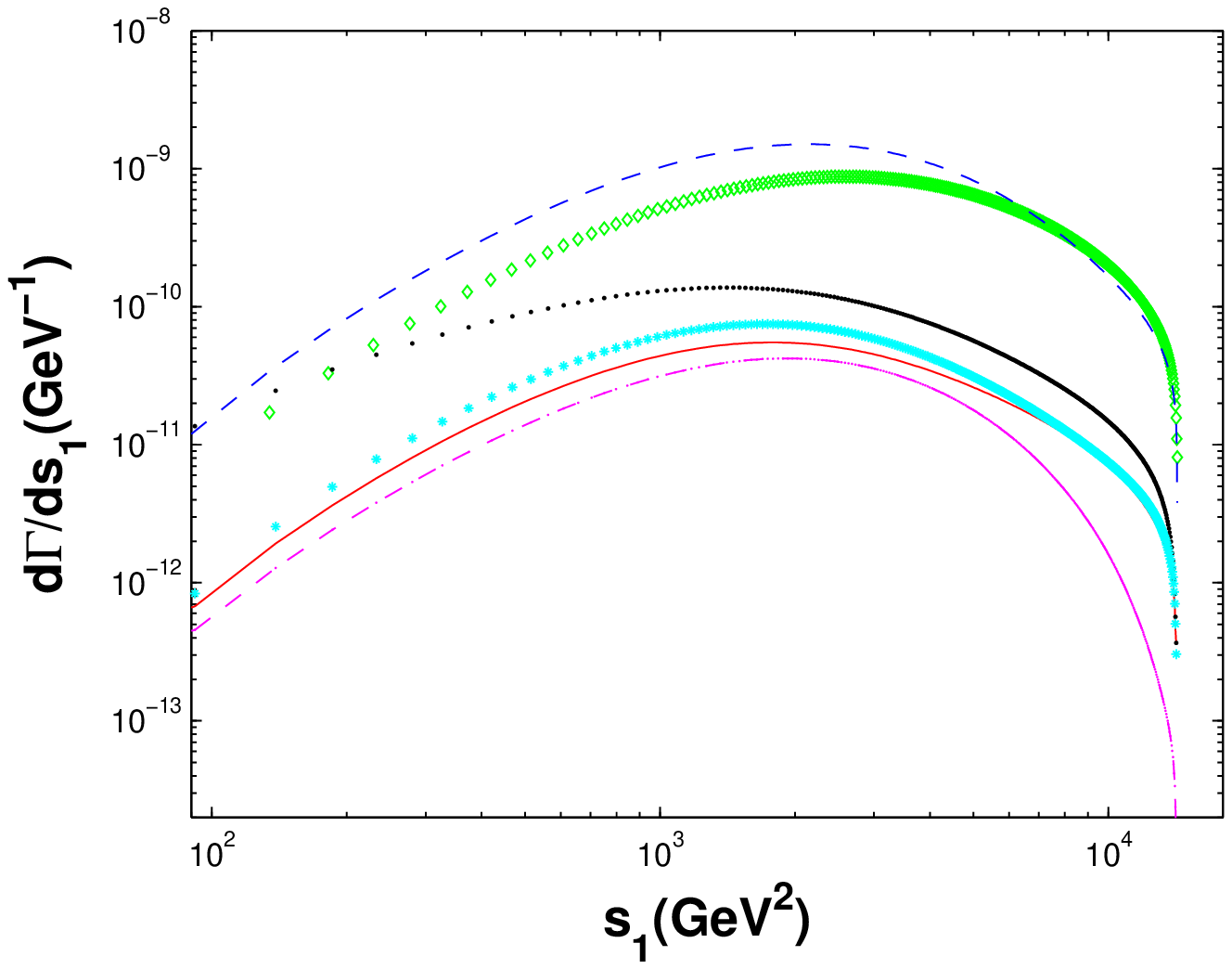}
\includegraphics[width=0.40\textwidth]{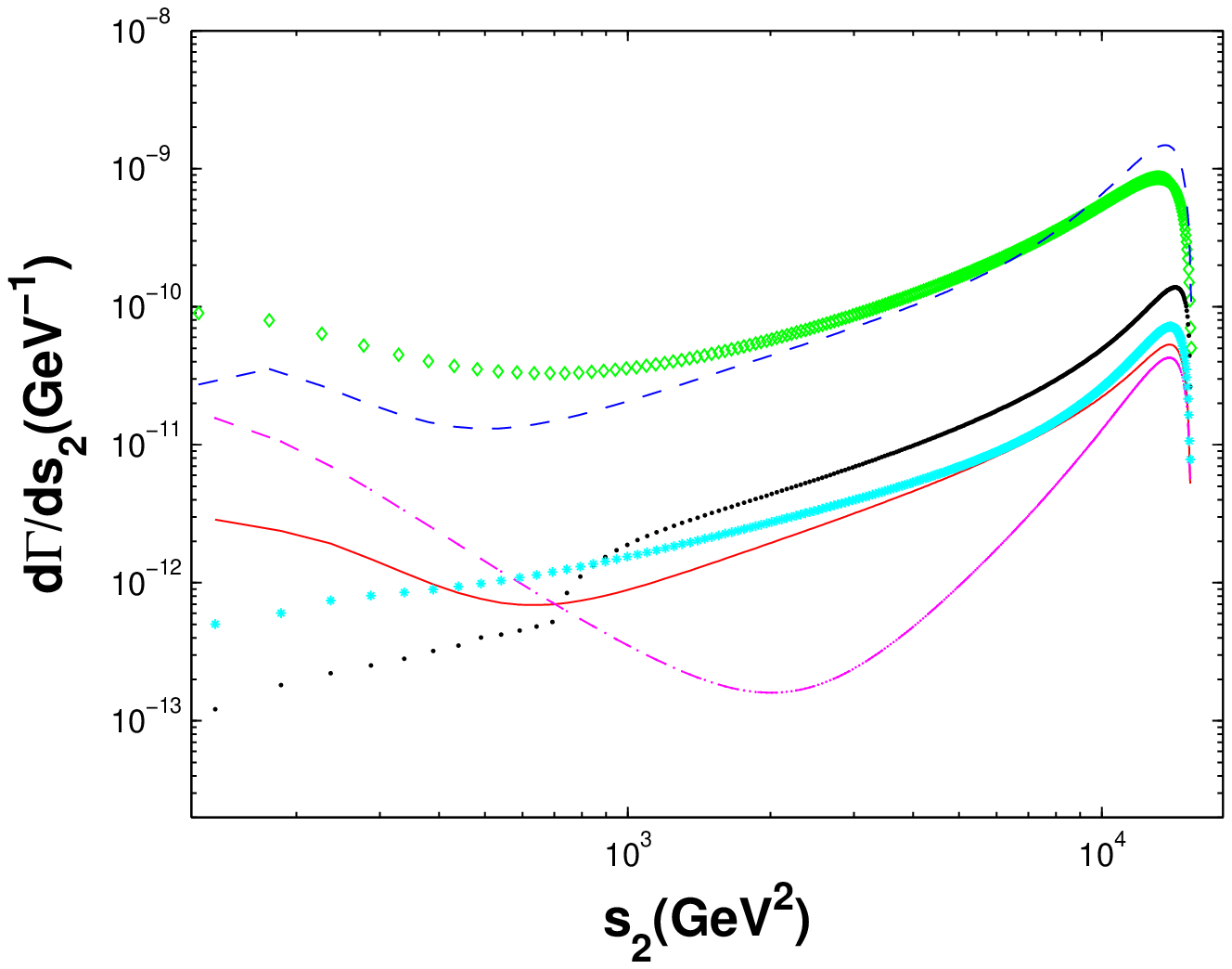}
\caption{ Differential decay widths $d\Gamma/ds_1$ and $d\Gamma/ds_2$ for $ H^{0}\rightarrow |(c\bar{b})[n]\rangle +b\bar{c}$, where the diamond line, the dashed line, the solid line, the dash-dotted line, the dotted line and the crosses line are for $|(c\bar{b})[1^1S_0]\rangle$, $|(c\bar{b})[1^3S_1]\rangle$, $|(c\bar{b})[1^1P_1]\rangle$, $|(c\bar{b})[1^3P_0]\rangle$, $|(c\bar{b})[1^3P_1]\rangle$, and $|(c\bar{b})[1^3P_2]\rangle$, respectively.} \label{H(bC)sds1ds2}
\end{figure}

\begin{figure}
\includegraphics[width=0.40\textwidth]{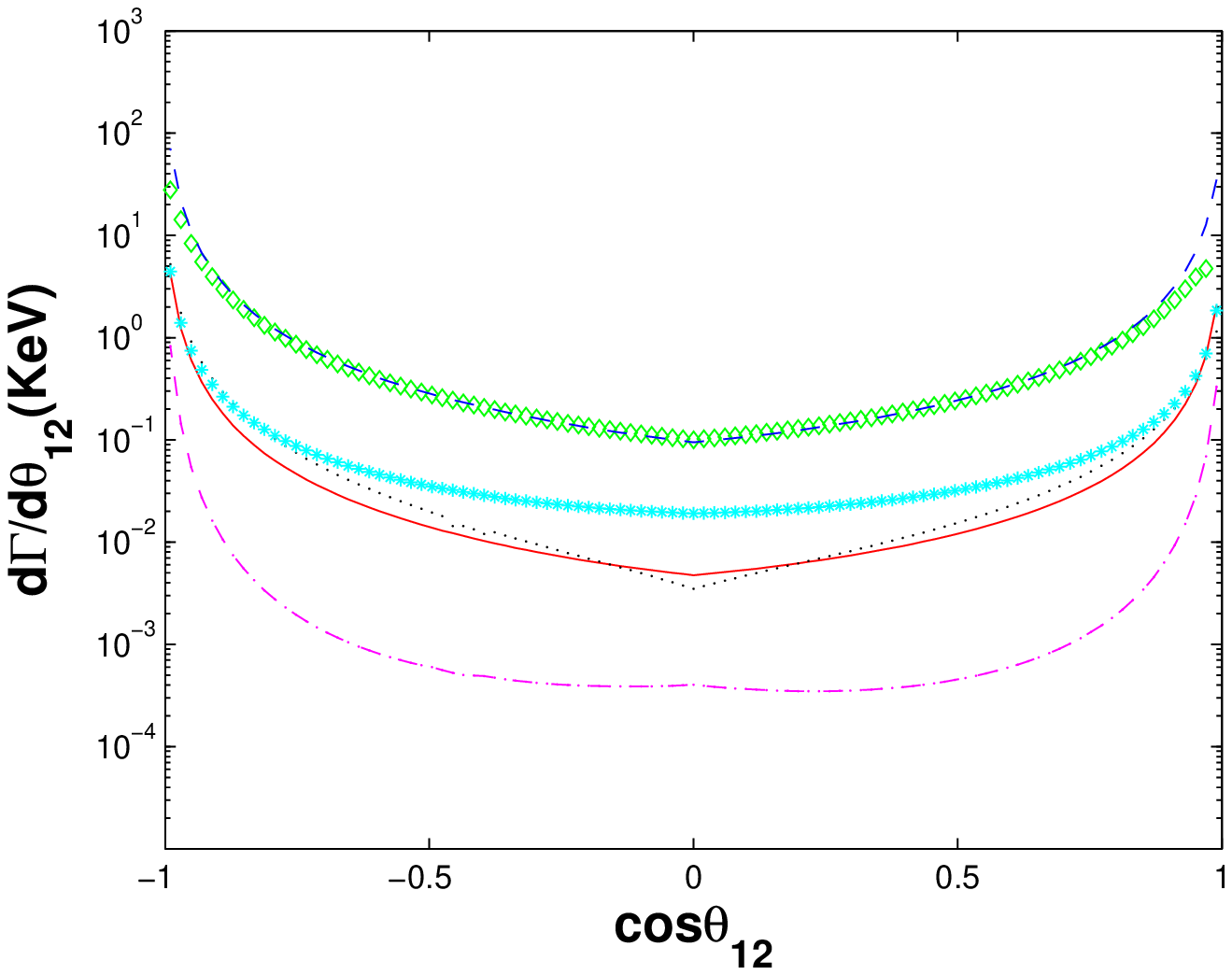}
\includegraphics[width=0.40\textwidth]{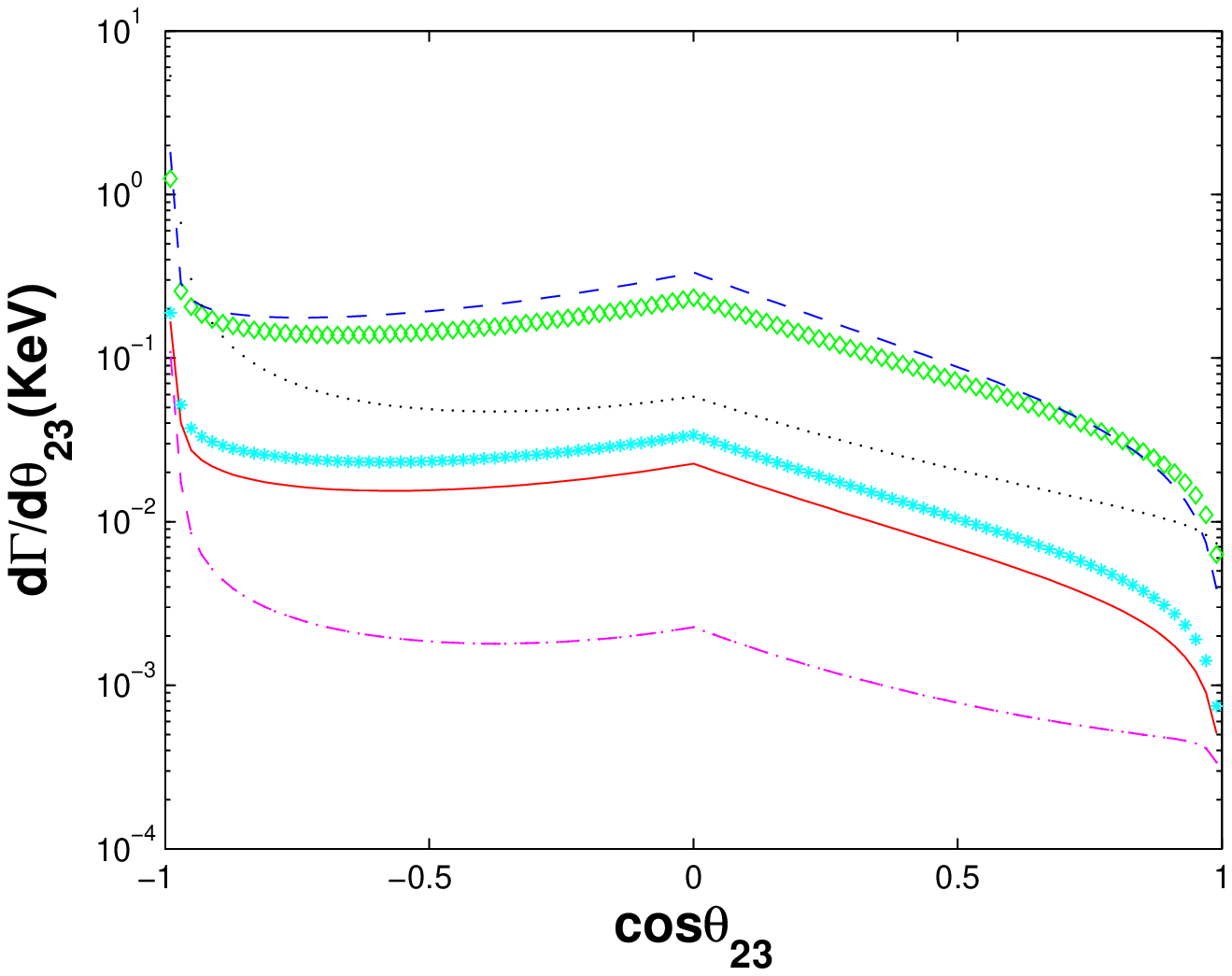}
\caption{ Differential decay widths $d\Gamma/dcos\theta_{12}$ and $d\Gamma/dcos\theta_{23}$ for $ H^{0}\rightarrow |(c\bar{b})[n]\rangle +b\bar{c}$, where the diamond line, the dashed line, the solid line, the dash-dotted line, the dotted line and the crosses line are for $|(c\bar{b})[1^1S_0]\rangle$, $|(c\bar{b})[1^3S_1]\rangle$, $|(c\bar{b})[1^1P_1]\rangle$, $|(c\bar{b})[1^3P_0]\rangle$, $|(c\bar{b})[1^3P_1]\rangle$, and $|(c\bar{b})[1^3P_2]\rangle$,  respectively.} \label{HBcdcos123}
\end{figure}

From Tables~\ref{tabrpb}-\ref{tabrpd}, it is noted that, in addition to the ground $1S$-level states, the $P$-states of $|(Q\bar{Q'})[n]\rangle$-quarkonium  can also provide sizable contributions to the total decay widths. Running at center-of-mass energy $\sqrt{s}=14TeV$, cross section of the Higgs boson production at the LHC is about $55 pb$. Given that the integrated luminosity is $3 ab^{-1}$, the HL-LHC would be produced $1.65\times10^8$ Higgs boson events per year~\cite{lhc}.

\begin{itemize}
\item For $|(c\bar{b})[n]\rangle$-quarkonium production via $H^0$ boson semiexclusive decays, the decay widths for $^1P_1$, $^3P_0$, $^3P_1$, and $^3P_2$-wave states is $4.81\%$ ($3.51\%$), $3.20\%$ ($2.34\%$), $11.69\%$ ($8.54\%$), and $6.14\%$ ($4.48\%$) of the decay width of the $|(c\bar{b})_{\textbf{1}}[^1S_0]\rangle$ ($|(c\bar{b})_{\textbf{1}}[^3S_1]\rangle$). Given that the integrated luminosity is $3 ab^{-1}$ and Running at center-of-mass energy $\sqrt{s}=14TeV$ at the HL/HE-LHC, cross section of Higgs boson production at the LHC is about $55 pb$. Then $2.26\times10^5$  $|(c\bar{b})_{\textbf{1}}[^1S_0]\rangle$, $3.08\times10^5$ $|(c\bar{b})_{\textbf{1}}[^3S_1]\rangle$, and $5.83\times10^4$ $|(c\bar{b})[1P]\rangle$- quarkonium events per year can be obtained at the HL/HE-LHC, where $1P$ is presented the summed decays widths of $|(c\bar{b})[^1P_1]\rangle$ and $|(c\bar{b})[^3P_J]\rangle$ ($J=0, 1, 2)$.
\end{itemize}

\begin{itemize}
\item For $|(c\bar{c})[n]\rangle$-quarkonium production via $H^0$ boson semiexclusive decays, the decay widths for $^1P_1$, $^3P_0$, $^3P_1$, and $^3P_2$-wave states is $10.80\%$ ($11.21\%$), $15.76\%$ ($16.34\%$), $6.86\%$ ($7.11\%$), and $7.12\%$ ($7.38\%$) of the decay width of the $|(c\bar{c})_{\textbf{1}}[^1S_0]\rangle$ ($|(c\bar{c})_{\textbf{1}}[^3S_1]\rangle$). $2.54\times10^4$  $|(c\bar{c})_{\textbf{1}}[^1S_0]\rangle$, $2.46\times10^4$ $|(b\bar{c})_{\textbf{1}}[^3S_1]\rangle$, and $1.03\times10^4$ $|(b\bar{c})[1P]\rangle$-quarkonium events per year can be obtained.
\end{itemize}

\begin{itemize}
\item For $|(b\bar{b})[n]\rangle$-quarkonium production via $H^0$ boson semiexclusive decays, the decay widths for $^1P_1$, $^3P_0$, $^3P_1$, and $^3P_2$-wave states is $3.14\%$ ($4.16\%$), $5.85\%$ ($7.76\%$), $5.69\%$ ($7.54\%$), and $2.30\%$ ($3.04\%$) of the decay width of the $|(b\bar{b})_{\textbf{1}}[^1S_0]\rangle$ ($|(b\bar{b})_{\textbf{1}}[^3S_1]\rangle$). At the LHC, $2.67\times10^4$  $|(b\bar{b})_{\textbf{1}}[^1S_0]\rangle$, $2.01\times10^4$ $|(b\bar{b})_{\textbf{1}}[^3S_1]\rangle$, and $4.54\times10^3$ $|(b\bar{b})[1P]\rangle$- quarkonium events per year can be obtained.
\end{itemize}

To better illustrate the relative importance of different production channels, we present the differential distributions
$d\Gamma/ds_{1}$, $d\Gamma/ds_{2}$, $d\Gamma/dcos\theta_{12}$, and $d\Gamma/dcos\theta_{23}$ for the $H^0\to |(c\bar{b})[n]\rangle +\bar{c}b$ processes in Figs. \ref{H(bC)sds1ds2},  \ref{HBcdcos123}. These figures show explicitly that the excited Fock states can provide sizable contributions in comparison to the lower Fock state $|(c\bar{b})[^1S_0]\rangle$ or $|(c\bar{b})[^3S_1]\rangle$ in almost the entire kinematical region.

As all the excited states decay to the ground state $|(Q\bar{Q})[^1S_0]\rangle$ with $100\%$ efficiency via electromagnetic or hadronic interactions, we can obtain the total decay width of $H^0$ boson decay channels within the B.T. potential:
\begin{eqnarray}
\Gamma{(H^0\to |(c\bar{b})[1^1S_0]\rangle +\bar{c}b)} &=&15.14\;{\rm KeV} \label{cc},\\
\Gamma{(H^0\to |(c\bar{c})[1^1S_0]\rangle +\bar{c}c)} &=&1.547\;{\rm KeV} \label{bc},\\
\Gamma{(H^0\to |(b\bar{b})[1^1S_0]\rangle +\bar{b}b)} &=&1.311\;{\rm KeV} \label{bb}.
\end{eqnarray}
where $v^2$=$0.20$, $0.30$, $0.10$ for the color-octet $|(c\bar{b})g[^1S_0]_{\textbf{8}}\rangle$ ($|(c\bar{b})g[^3S_1]_{\textbf{8}}\rangle$), $|(c\bar{c})g[^1S_0]_{\textbf{8}} \rangle$ ($|(c\bar{c})g[^3S_1]_{\textbf{8}}\rangle$), and $|(b\bar{b})g[^1S_0]_{\textbf{8}} \rangle$ ($|(b\bar{b})g[^3S_1]_{\textbf{8}} \rangle$) are adopted, respectively.

Running at the center-of-mass energy $\sqrt{S}=14$ TeV at the HL/HE-LHC \cite{eml,lhc} and with luminosity $10^{34} cm^{-2} s^{-1}$, one may expect to produce about $1.65\times10^{8}$ $H^0$ boson per year. Then we can estimate the event number of $|(Q\bar{Q'})\rangle$-quarkonium production through $H^0$ boson decays, i.e., about $5.6\times10^5$ $|(c\bar{b})[n]\rangle$ (or $|(b\bar{c})[n]\rangle$)-quarkonium events, $4.7\times10^4$ $|(c\bar{c})[n]\rangle$-quarkonium events, $4.9\times10^4$ $|(b\bar{b})[n]\rangle$-quarkonium events per year.  The upgrade HE/HL-LHC and the newly purposed $H^0$ factory with luminosity $10^{36} cm^{-2} s^{-1}$, the possibility to study $|(Q\bar{Q'})[n]\rangle$-quarkonium via $H^0$ boson decays at the SPPC, ILC, and CEPC is worth serious consideration.

\subsection{Uncertainty analysis}

In the subsection, we discuss the uncertainties for the $|(c\bar{b})[n]\rangle$-quarkonium production through Higgs boson decays. For the present calculation, their main uncertainty sources include  the renormalization scale $\mu_R$, the nonperturbative bound state matrix elements, and the constituent quark masses $m_c$ and $m_b$. These parameters are the main uncertainty source for estimating the $|(c\bar{b})[n]\rangle$-quarkonium production. Here we only discuss the decay widths of the $|(c\bar{b})[n]\rangle$-quarkonium production via $H^0$ decays under varying the constituent quark masses of the $|(c\bar{b})[n]\rangle$-quarkonium. In the following, we shall concentrate our attention on the uncertainties caused by $m_c$ and $m_b$, whose center values are taken as $m_c=1.45\pm0.10$ GeV and $m_b=4.85\pm0.20$ GeV for $S$-states, and $m_c=1.75\pm0.10$ GeV and $m_b=4.93\pm0.20$ GeV for $P$-states. And for clarity, when discussing the uncertainty caused by one parameter, the other parameters are fixed to be their center values.

In the Tables.~\ref{tabrpe} and~\ref{tabrpf}, it can be found that sizable uncertainties for varying $m_c$ and $m_b$. The decay width will decrease with the increment mass of $m_c$. But the decay width will increase with the increment mass of $m_b$.

\begin{table}
\caption{Uncertainties for the decay width of the processes $H^0\to |(c\bar{b})[n]\rangle +b\bar{c}$ under the B.T. potential ($n_f=3$)~\cite{lx, pot2}.}
\begin{tabular}{|c|c|c|c|c|c|}
\hline\hline
~~~&$m_c$~$(GeV)$~&~1.35~&~1.45~&~1.65~\\
S-states&$\Gamma(H^0\to |(c\bar{b})[^1S_0]_{\textbf{1}}\rangle)$({KeV})&7.108&5.736&4.697\\
~&$\Gamma(H^0\to |(c\bar{b})[^3S_1]_{\textbf{1}}\rangle)$({KeV})&9.981&7.857 &6.283\\
\hline\hline
~&$m_c$~$(GeV)$~&~1.65~&~1.75~&~1.85~\\
P-states&$\Gamma(H^0\to |(c\bar{b})[^1P_1]_{\textbf{1}}\rangle)$({KeV})&0.3763&0.2761&0.2145\\
~&$\Gamma(H^0\to |(c\bar{b})[^3P_0]_{\textbf{1}}\rangle)$({KeV})&0.2326&0.1838&0.1478\\
~&$\Gamma(H^0\to |(c\bar{b})[^3P_1]_{\textbf{1}}\rangle)$({KeV})&0.8879&0.6706&0.5143\\
~&$\Gamma(H^0\to |(c\bar{b})[^3P_2]_{\textbf{1}}\rangle)$({KeV})&0.4820&0.3521&0.2615\\
\hline\hline
Color-octet&$m_c$~$(GeV)$~&~1.35~&~1.45~&~1.65~\\
S-states&$\Gamma(H^0\to |(c\bar{b})g[^1S_0]_{\textbf{8}}\rangle)$({KeV})&0.889$v^{4}$&0.717$v^{4}$&0.587$v^{4}$\\
~&$\Gamma(H^0\to |(c\bar{b})g[^3S_1]_{\textbf{8}}\rangle)$({KeV})&1.248$v^{4}$&0.982$v^{4}$&0.785$v^{4}$\\
\hline\hline
\end{tabular}
\label{tabrpe}
\end{table}

\begin{table}
\caption{Uncertainties for the decay width of the processes $H^0\to |(c\bar{b})[n]\rangle +b\bar{c}$ under the B.T. potential ($n_f=3$)~\cite{lx, pot2}.}
\begin{tabular}{|c|c|c|c|c|c|}
\hline\hline
~~~~&~~$m_b$~$(GeV)$&~4.65~&~4.85~&~5.05~\\
S-states&$\Gamma(H^0\to |(c\bar{b})[^1S_0]_{\textbf{1}}\rangle)$({KeV})&5.280&5.736&6.210\\
~&$\Gamma(H^0\to |(c\bar{b})[^3S_1]_{\textbf{1}}\rangle)$({KeV})&7.141&7.857&8.611\\
\hline\hline
~&$m_b$~$(GeV)$~&~4.73~&~4.93~&~5.13~\\
P-states&$\Gamma(H^0\to |(c\bar{b})[^1P_1]_{\textbf{1}}\rangle)$({KeV})&0.2614&0.2761&0.3016\\
~&$\Gamma(H^0\to |(c\bar{b})[^3P_0]_{\textbf{1}}\rangle)$({KeV})&0.1771&0.1838&0.1907\\
~&$\Gamma(H^0\to |(c\bar{b})[^3P_1]_{\textbf{1}}\rangle)$({KeV})&0.6244&0.6706&0.7175\\
~&$\Gamma(H^0\to |(c\bar{b})[^3P_2]_{\textbf{1}}\rangle)$({KeV})&0.3209&0.3521&0.3845\\
\hline\hline
Color-octet&~~$m_b$~$(GeV)$&~4.65~&~4.85~&~5.05~\\
S-states&$\Gamma(H^0\to |(c\bar{b})g[^1S_0]_{\textbf{8}}\rangle)$({KeV})&0.660$v^{4}$&0.717$v^{4}$&0.776$v^{4}$\\
~&$\Gamma(H^0\to |(c\bar{b})g[^3S_1]_{\textbf{8}}\rangle)$({KeV})&0.893$v^{4}$&0.982$v^{4}$&1.076$v^{4}$\\
\hline\hline
\end{tabular}
\label{tabrpf}
\end{table}

Adding all the uncertainties caused by the constituent quark masses $m_c=1.45\pm0.10$ GeV and $m_b=4.85\pm0.20$ GeV for $S$-states, and $m_c=1.75\pm0.10$ GeV and $m_b=4.93\pm0.20$ GeV for $P$-states in quadrature for the process $H^0\to |(c\bar{b})[n]\rangle+b\bar{c}$, we can obtain

\begin{eqnarray}
\Gamma{(H^0\to |(c\bar{b})[1^1S_0]_{\textbf{1}}\rangle+b\bar{c}}&=&5.736^{+1.452}_{-1.135}\;{\rm KeV},\nonumber\\
\Gamma{(H^0\to |(c\bar{b})[1^3S_1]_{\textbf{1}}\rangle+b\bar{c}}&=&7.857^{+2.254}_{-1.729}\;{\rm KeV},\nonumber\\
\Gamma{H^0\to |(c\bar{b})[1P]_{\textbf{1}}\rangle+b\bar{c}}~~&=&1.483^{+0.508}_{-0.359}\;{\rm KeV}.
\end{eqnarray}

If the excited $|(c\bar{b})[n]\rangle$-quarkonium states decay to the ground spin-singlet $S$-wave state $|(c\bar{b})[1^1S_0]\rangle$ with $100\%$ efficiency via hadronic interactions or electromagnetic, we can obtain the total decay width of the Higgs boson decay channels under the B.T. potential.
\begin{eqnarray}
\Gamma{(H^0\to |(c\bar{b})[1^1S_0]\rangle+b\bar{c})}&=&15.14^{+2.735}_{-2.105}\;{\rm KeV} \label{tWbs1}.
\end{eqnarray}

\section{Conclusions}

In this paper, we have made a detailed study on the $|(c\bar{b})[n]\rangle$ (or $|(b\bar{c})[n]\rangle$), $|(c\bar{c})[n]\rangle$, and $|(b\bar{b})[n]\rangle$-quarkonium via $H^0$ boson semiexclusive decays under the NRQCD framework, i.e., $H^0\to |(b\bar{c})[n]\rangle +\bar{b}c$ (or $H^0\to |(c\bar{b})[n]\rangle +\bar{c}b$), $H^0\to |(c\bar{c})[n]\rangle +\bar{c}c$, and $H^0\to |(b\bar{b})[n]\rangle +\bar{b}b$, where $[n]$ stands for $[^1S_0]$, $[^3S_1]$, $[^1P_1]$, and $[^3P_J]$, ($J=[0, 1, 2]$). To provide the analytical expressions as simply as possible, we have adopted the improved trace technology to derive Lorentz-invariant expressions for $H^0$ boson decay processes at the amplitude level. Such a calculation technology will be very helpful for dealing with processes with massive spinors.

Numerical results show that $P$-states of $|(Q\bar{Q'})[n]\rangle$-quarkonium in addition to the ground $1S$ wave states can also provide sizable contributions to the heavy quarkonium production through $H^0$ boson decays, so one needs to take the excited wave states into consideration for a sound estimation. If all of the excited heavy quarkonium $1P$ Fock states almost decay to the ground spin-singlet $S$ wave state $|(Q\bar{Q'})[1^1S_0]\rangle$ via electromagnetic or hadronic interactions, we obtain the total decay width the total decay width for $|(Q\bar{Q'})\rangle$-quarkonium production through $H^0$ boson decays as shown by Eqs.~(\ref{cc})-(\ref{bb}). At the LHC at running with center-of-mass energy $\sqrt{S}=14$ TeV with the luminosity ${\cal L}\propto 10^{34}cm^{-2}s^{-1}$, due to the high collision energy and high luminosity, sizable heavy quarkonium events can be produced through $H^0$ boson decays; i.e., about $5.6~\times10^{5}$ of $(c\bar{b})$ (or $(b\bar{c})$) meson,  $4.7~\times10^{4}$ of $(c\bar{c})$ meson, and $4.9~\times10^{4}$ of $(b\bar{b})$ meson events per year can be obtained. At the newly purposed $H$ factory with the high luminosity ${\cal L}\propto 10^{36}cm^{-2}s^{-1}$, the $|(Q\bar{Q'})\rangle$-quarkonium through $H^0$ boson decays will be more abundantly produced. Therefore, one needs to take these $P$-states into consideration for a sound evaluation.

\hspace{2cm}

{\bf Acknowledgements}: We thank Professor Xing-Gang Wu for useful discussions.

\end{document}